\documentclass[twocolumn,prb,aps,showpacs,superscriptaddress]{revtex4-1}
\usepackage{graphicx}
\usepackage{color}
\usepackage{dcolumn}
\usepackage{amsmath}
\usepackage{physics}
\usepackage{tabularx}
\usepackage{bm}
\usepackage{lineno}

\newlength{\figwidth} 
 \newlength{\figwidthb} 
\newcommand{\spin}{$\mathbf{S}$}
\newcommand{\orb}{$\mathbf{L}$}
\newcommand{\mom}{$\mathbf{M}$}

\newcommand{\q}{$\mathbf{q}$}

\begin{document}

\title{Resonant inelastic x-ray scattering operators for $t_{2g}$ orbital 
systems}

\author{B. J. Kim}
\affiliation{Max Planck Institute for Solid State Research, 
Heisenbergstrasse 1, D-70569 Stuttgart, Germany} 
\affiliation{Department of Physics, Pohang University of Science 
and Technology, Pohang 790-784, Republic of Korea}
\affiliation{Center for Artificial Low Dimensional Electronic Systems, Institute for Basic Science (IBS), 77 Cheongam-Ro, Pohang 790-784, Republic of Korea}
\author{Giniyat Khaliullin}
\affiliation{Max Planck Institute for Solid State Research, 
Heisenbergstrasse 1, D-70569 Stuttgart, Germany}

\date{\today}

\begin{abstract}
We derive general expressions for resonant inelastic x-ray scattering (RIXS)
operators for $t_{2g}$ orbital systems, which exhibit a rich array of
unconventional magnetism arising from unquenched orbital moments. 
Within the fast collision approximation, which is valid especially for 4$d$ 
and 5$d$ transition metal compounds with short core-hole lifetimes, the RIXS
operators are expressed in terms of total spin and orbital angular momenta of
the constituent ions. We then map these operators onto pseudospins that 
represent spin-orbit entangled magnetic moments in systems with strong 
spin-orbit coupling. Applications of our theory to such systems as iridates 
and ruthenates are discussed, with a particular focus on compounds based 
on $d^4$ ions with Van Vleck-type nonmagnetic ground state.  

\end{abstract}


\maketitle
\section{Introduction}
Raman scattering of photons in the infrared and visible range by a quantum of
magnetic excitation, or magnon, was observed and understood by the late
1960s\cite{FL}. Corresponding advances in the x-ray regime\cite{AmentReview}
have only been achieved recently thanks to advances in x-ray technologies,
including intense sources from modern synchrotrons, high resolution and
efficiency optics, and multi-channel detectors. X-rays can transfer momenta
($\mathbf{q}$) of the order of  reciprocal-lattice spacings, which is a
significant advantage over the Raman light scattering which is virtually
limited to $\mathbf{q}$\,=\,0 modes.  

In 2010, Braicovich {\it et al.}~made the first observation of dispersive
single-magnon excitations using soft x-rays (Cu $L$ edge\,$\sim$\,930 eV) on a
thin film of La$_2$CuO$_4$ (Ref.~\citenum{BraicovichPRL10}), which was shortly
followed by Kim {\it et al.} who used hard x-rays (Ir $L_3$ edge\,$\sim$\,11.2
keV) on a single crystal of Sr$_2$IrO$_4$ (Ref.~\citenum{KimPRL12}). These
materials, with their large magnon energy scales, have served as ideal systems
to explore magnetic scattering in the 
early development stage of Raman x-ray scattering, which is now more commonly
known as resonant inelastic x-ray scattering (RIXS). Over the past years, RIXS
has rapidly become a complementary tool to inelastic neutron scattering (INS)
for studies of magnetic materials, and has witnessed a dramatic enhancement in
its energy resolution, heading toward sub-10 meV resolution\cite{Kim17}. 

RIXS has broad sensitivity to charge, orbital, spin, and lattice degrees of
freedom in a solid, and in general probes different quantities as compared to
INS. For magnetic insulators, however, magnetic and charge scatterings can
usually be separated by their different energy scales, and the spectra at
energies below the charge gap are dominated by magnetic scattering. In
particular, when the orbital moment can be approximated as fully quenched, as
for example in $S$=1/2 cuprates and $S$=1 nickelates, the RIXS cross section
reduces to usual spin-spin correlation 
functions\cite{AmentPRL09,HaverkortPRL10}.  

However, many transition-metal (TM) compounds possess unquenched orbital
degrees of freedom active at low energies\cite{Khaliullin05}. For example,
dispersive orbital excitations have been observed by RIXS in Mott insulating
titanates\cite{Ulr09}, vanadates\cite{Ben13}, manganites\cite{Web10}, and
iridates\cite{KimPRL12,Kim14,YJK17}, and described
theoretically\cite{KimPRL12,AmentPRB09,Kim14}. Nevertheless, RIXS cross 
sections for orbitally active systems still lack a general theoretical 
framework, which is particularly important for the emerging class
of 4$d$ and 5$d$ transition-metal compounds with strong spin-orbit 
coupling (SOC). RIXS is particularly well matched to 5$d$ TM compounds, 
because (i) the x-ray optics to implement RIXS is relatively 
straightforward\cite{Gog13}, (ii) wavelength at
5$d$ TM $L$ edges ($\sim$1$\textrm{\AA}$) is small enough to cover many
Brillouin zones (note that in RIXS there is no suppression of magnetic
scattering at high \q\ due to the form factor as in INS), (iii) and thus
equivalent $\mathbf{q}$-points can be measured in different scattering
geometries, allowing differentiation among modes of different symmetries.  

The aim of this paper is to provide general expressions for RIXS cross
sections for magnetic insulators in which both spin and orbital degrees of
freedom are active and reside in the orbitals of $t_{2g}$ symmetry. This is
the case for many compounds with TM ions in the octahedral coordination
geometry, which allows the local symmetry around the TM ion to be approximated
as a small deviation from the cubic limit. Within the fast collision 
approximation\cite{VeePRL06}, the RIXS operators are expressed in terms of 
total orbital and spin angular momentum operators active at low-energies. 
These operators depend only on the electron occupation number 
($d^n$;\,n\,=\,1\,--\,5), symmetry of the probe, and the resonant 
TM edge ($L_2$ and $L_3$).   

The paper is organized as follows. In Sec.~II we start with a brief review of
the ``direct'' RIXS process sensitive to both single- and double-magnon
excitations, and derive from it a set of general expressions for RIXS
scattering operators. We shall not be concerned with the indirect RIXS
process, which is generally insensitive to single magnon 
excitations\cite{AmentReview}. In Sec.~III, we map the RIXS operators onto the
pseudospins representing spin-orbit entangled magnetic moments in the strong 
SOC limit, with applications to iridates and ruthenates in mind. The effect 
of symmetry-lowering lattice distortions will be discussed. Sec.~IV concludes 
the paper with a brief summary and an outlook. 

\section{General expressions for RIXS operators}

\begin{figure}
\centerline{\includegraphics[width=1\columnwidth,angle=0]{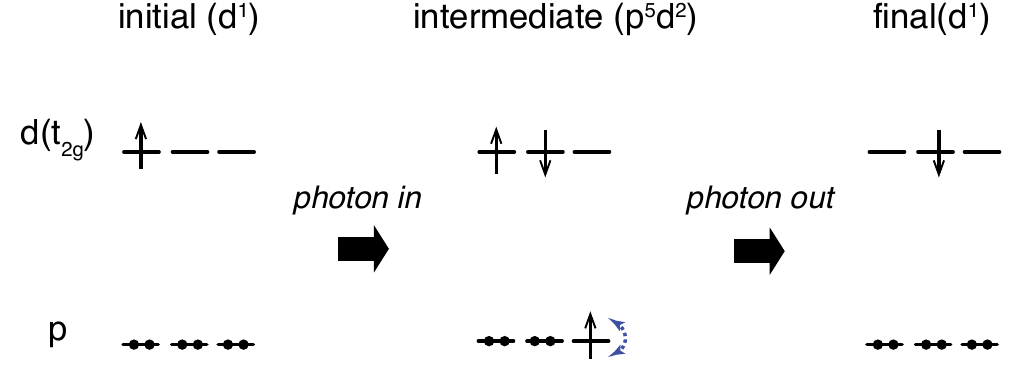}}
\caption{A schematic of direct RIXS process at TM L edges. For concreteness we
  illustrate the case for $d^1$ system. A photon is absorbed exciting a $p$
  core electron to the $d$ valence level through a dipole transition. The
  intermediate states have one extra electron in the $d$ manifold and a hole
  in the $p$ manifold. The core $p$ level spin is not conserved because of
  strong SOC. The intermediate states decay back to one of the $d^{n}$ 
  multiplets by emitting a photon and completes the RIXS process.} 
\end{figure}

Formally, the RIXS process for magnetic excitations is identical to that of
Raman scattering. It involves a radiative excitation to a set of intermediate
states, and a subsequent de-excitation to a final state, which can be
different from the initial state if there is a non-zero energy transfer. The
initial and final states can have different spin quantum numbers ($S_z$) if
there is sufficiently strong SOC in the intermediate states, even though
dipole transitions themselves conserve $S_z$. Figure 1 describes a typical
RIXS process at the $L$ edges of TM compounds involving dipole transitions
between core $p$ and valence $d$ states, given by the operator  

\begin{align}
D&=D(t_{2g})+D(e_{g})\nonumber\\
&\propto\epsilon_x [(d_{zx}^\dagger p_z+d_{xy}^\dagger p_y) 
+ \tfrac{2}{\sqrt{3}} d_{x^2}^\dagger p_x] \nonumber \\
 &+\epsilon_y [(d_{xy}^\dagger p_x+d_{yz}^\dagger p_z) 
+ \tfrac{2}{\sqrt{3}} d_{y^2}^\dagger p_y] \nonumber \\
 &+\epsilon_z [(d_{yz}^\dagger p_y+d_{zx}^\dagger p_x) 
+ \tfrac{2}{\sqrt{3}} d_{z^2}^\dagger p_z]. 
\end{align}
\noindent
Here, $\bm{\epsilon}$ denotes the x-ray polarization, and $d$ and $p$
annihilate an electron in the respective orbitals. The $d$-orbitals are
divided into two sets of $t_{2g}$ and $e_g$ symmetries, and shorthand notations 
$d_{z^2}$\,$\equiv$\,$d_{3z^2-r^2}$ and 
$d_{x^2/y^2}$\,=\,$-\tfrac{1}{2}d_{z^2} \pm\tfrac{\sqrt 3}{2} d_{x^2-y^2}$ are used
for the latter. Summation over spin projection is implied. 
In a crystal field 
of octahedral symmetry, the $t_{2g}$ and $e_g$ levels split by a large energy 
$10Dq$ of the order of 2-4 eV (increasing as one goes from 3$d$ to 5$d$ ions). 
Since $10Dq$ splitting is an order of value larger than SOC constant 
$\zeta_d\sim 0.1$-$0.4$~eV for $d$ electrons, the $t_{2g}$ and $e_g$ orbital 
basis used in Eq.~(1) is most natural and convenient in practice. In contrast, 
for the core $p$ states with strong SOC, a total angular momentum 
representation will be used in calculations. 

Because the intermediate states are not detected, scattering amplitudes 
through all possible intermediate states ($\ket{m}$) add up coherently 
weighted by their different energies ($E_m$). Thus, the RIXS operator is 
expressed as 
\begin{equation}
R=\sum_m \frac{D^\dagger \ket{m}\bra{m} D}{E-E_m+i\Gamma/2}\;,
\end{equation}
where $E$ and $\Gamma$ denote the incident x-ray energy and lifetime (full
width half maximum) of the core hole, respectively. The scattering amplitude
is maximized when $E$\,$\approx$\,$E_m$, which defines the resonant condition.  

Since the energy separation between the $L_3$ and $L_2$ edges of 4$d$ (5$d$)
TM ions are of the order $\sim$100 eV ($\sim$1 keV), much larger than
$\Gamma$, resonances at the two $L$ edges can be taken as two separate RIXS
processes to an excellent approximation. Thus, \{$m$\} is divided into two
subsets according to the $p$ core-hole total angular momentum $J$,
\{$\bar{p}_{1/2}d^{n+1}$\} and \{$\bar{p}_{3/2}d^{n+1}$\}, leading to two
distinct RIXS operators for the two resonant edges, $L_2$ and $L_3$
respectively. 

The complex time dynamics of the intermediate states makes the RIXS process 
hard to analyze microscopically. However, as far as one is concerned with 
the low-energy excitations in Mott insulators, the problem of the 
fast-evolving intermediate states can be disentangled and cast in the form of 
frequency independent constants\cite{AmentPRB09,HaverkortPRL10,Sav15}. This 
results in an effective RIXS operator formulated in terms of low-energy spin 
and orbital degrees of freedom. 

Although this ``fast collision'' approximation may be questionable for 3$d$ TM
compounds with relatively shallow core levels, especially for doped systems 
where d-electron time scales become comparable to that of the intermediate 
states\cite{DemlerPRL14}, it is well justified for 4$d$ and 5$d$ TM compounds 
with $\Gamma$ typically of the order of several eV (Ref.~\citenum{Krause}), 
i.e. much larger than spin-orbital energy scales in Mott insulators. 

Within this  approach, the RIXS operator is approximated as 
$R\propto D^\dagger(\epsilon')D(\epsilon)$. Note that $R$ depends on two
photon polarizations, $\bm{\epsilon}$ (incoming) and $\bm{\epsilon'}$ 
(outgoing), the product of which can be decomposed into symmetric and 
antisymmetric combinations. As a result, $R$ operator is decomposed as
\begin{equation}
R\propto D^\dagger D=\frac{1}{3}(R_Q+iR_M),
\end{equation}
where $R_Q$ and $R_M$ describe the quadrupolar and dipolar RIXS channels, 
respectively. $R_Q$ can further be decomposed into diagonal and off-diagonal 
components, which couple to photon polarizations as: 
\begin{eqnarray}
R_Q&=& \sum_{\alpha}\epsilon_\alpha\epsilon'_\alpha Q_{\alpha\alpha}-
\tfrac{1}{2}\sum_{\alpha>\beta}(\epsilon_\alpha\epsilon'_\beta+
\epsilon_\beta\epsilon'_\alpha)\,Q_{\alpha\beta}\;, \\
R_M&=&\tfrac{1}{2}(\bm{\epsilon}\times\bm{\epsilon'})\cdot{\mathbf N}\;. 
\end{eqnarray}
To obtain explicit formulae for the quadrupolar tensor $Q_{\alpha\beta}$
and the magnetic vector ${\mathbf N}$ operators in the above equations, one 
has to ({\it i}) eliminate the core $p$-holes in the product $D^\dagger D$, and 
({\it ii}) express transitions within the $d$-shell in terms of total spin 
${\bf S}$ and effective orbital ${\bf L}$ angular momenta of multi-electron 
$d(t_{2g}^n)$ configuration. While the step ({\it i}) is trivial (projecting 
$p_{x,y,z}$ states onto $p_{1/2}$ and $p_{3/2}$ manifolds, separating $L_2$
and $L_3$ scattering channels), the step ({\it ii}) deserves more explanations 
as follows. 

First, while the above equations are completely general (provided that ``fast
collision'' picture is valid), here we confine ourselves to systems based 
on $d^n$-ions with $n=1,...,5$. Under strong octahedral field $10Dq$, the 
dominant electron configuration is then $t_{2g}^n$, with only small admixture 
of higher-lying $t_{2g}^{n-1}e_g$ states due to SOC, covalency,
etc.\cite{Tho68}. This admixture results in corrections of the order of 
$\zeta_d/10Dq \sim 0.1$ which we neglect in our calculations of matrix 
elements. Second, we focus on low-energy collective excitations within the 
$t_{2g}^n$ manifold, leaving aside high-energy ($10Dq$\,$\sim$\,2-4 eV) local 
transitions from $t_{2g}$ to empty $e_g$ levels (which are observable both in 
$L_2$ and $L_3$ edges\cite{Mat16} but not of interest in the present context). 
In technical terms, the above conditions imply that the dipolar transitions 
to $e_g$ states in Eq.~(1) can be safely neglected, leaving us with the RIXS 
processes operating within the $t_{2g}^n$ spin-orbital subspace alone. Finally, 
we assume that Hund's coupling is strong enough to form a maximal spin $S$ 
allowed within the $t_{2g}^n$ configuration (i.e. $S$=1/2 for $n$=1 and 5, 
$S$=1 for $n$=2 and 4, and $S$=3/2 for $n$=3). Except for the pure-spin $n$\,=\,3 case, 
there remains three-fold orbital degeneracy of $t_{2g}^n$ multi-electron 
configuration which is conveniently described by an effective orbital moment 
$L=1$ (Ref. \citenum{Abragam}). A meaning of this mapping is that an octahedral crystal 
field separates the initially large Hilbert space of $d^n$ configuration into two 
subsets split by large $10Dq$ that much exceeds SOC, noncubic crystal fields, 
etc. As far as one is interested in low-energy physics within the $t_{2g}^n$ 
sector, the physical observables can be then concisely expressed via 
effective $L$ operator. In the context of RIXS problem, the task in 
step ({\it ii}) above is then to express various bilinear forms 
$d_{yz}d_{zx}^\dagger$, etc... (that appear in the product of 
$D^\dagger D$) in terms of total ${\bf S}$ and effective ${\bf L}$ momenta. 
The resulting RIXS operators cover all the magnetic and quadrupole transitions 
within $(2L+1)(2S+1)$ levels of $t_{2g}^n$ configuration split by SOC and 
noncubic crystal fields. 

After somewhat tedious but straightforward calculations, the $Q_{\alpha\beta}$ 
and ${\mathbf N}$ operators listed in Table I are obtained. This is the 
central result of this paper, which is used in the following sections.  

\begin{table*} 
\newcolumntype{R}{>{\raggedleft\arraybackslash}X}
	\begin{tabularx}{0.9\textwidth}{|l|c||c|}	
                \multicolumn{1}{c}{${Q_{zz}}$ (quadrupole)}& 
		\multicolumn{1}{c}{$L_3$ edge} &
		\multicolumn{1}{c}{$L_2$ edge}\\	 
		\cline{1-3}
		$~~~d^1, (-1)d^5$ &$-2L_z^2 + 2L_zS_z$ &  $-L_z^2 - 2L_zS_z$ \\
		\cline{1-3}
		$~~~d^2, (-1)d^4$ &$2L_z^2 + L_zS_z$ & $ L_z^2 - L_zS_z$ \\
		\cline{1-3}
		\multicolumn{3}{l}{}\\
		\multicolumn{3}{l}{}\\
		\multicolumn{1}{c}{${Q_{xy}}$ (quadrupole)}& 
		\multicolumn{1}{c}{$L_3$ edge} &
		\multicolumn{1}{c}{$L_2$ edge}\\
		\cline{1-3}
		$~~~d^1, (-1)d^5$ & $-2L_xL_y - 2L_yL_x +2L_xS_y + 2L_yS_x$ 
                             & $-L_xL_y - L_yL_x - 2L_xS_y - 2L_yS_x$\\
		\cline{1-3}
		$~~~d^2, (-1)d^4$ & $2L_xL_y + 2L_yL_x +L_xS_y + L_yS_x$ 
                            & $L_xL_y + L_yL_x - L_xS_y - L_yS_x$ \\
		\cline{1-3}
		\multicolumn{3}{l}{}\\
		\multicolumn{3}{l}{}\\
			\multicolumn{1}{l}{${N_{z}}$ (magnetic)}& 
			\multicolumn{1}{c}{$L_3$ edge} &
			\multicolumn{1}{c}{$L_2$ edge}\\
			\cline{1-3}
                        $~~~d^1, d^5$ & $2L_z -4S_z +8L_z^2S_z - 
2L_z({\bf L}\cdot{\bf S}) - 2({\bf L}\cdot{\bf S})L_z$ &  
$L_z +4S_z -8L_z^2S_z + 2L_z({\bf L}\cdot{\bf S}) + 2({\bf L}\cdot{\bf S})L_z$\\
			\cline{1-3}
			$~~~d^2$, $d^4$ & $2L_z - 4L_z^2S_z + 
L_z({\bf L}\cdot{\bf S}) + ({\bf L}\cdot{\bf S})L_z$ & 
$L_z + 4L_z^2S_z - L_z({\bf L}\cdot{\bf S}) - ({\bf L}\cdot{\bf S})L_z$ \\
			\cline{1-3}
			$~~~d^3$ &  $(4/3) S_z$ & $-(4/3) S_z$ \\
			\cline{1-3}
	\end{tabularx}

\caption{Quadrupolar and magnetic RIXS operators for $d^n$ systems at the
  $L_3$ and $L_2$ edges. Magnetic dipole and electric quadrupole operators 
  couple to antisymmetric 
  ($\epsilon_\alpha\epsilon_\beta'$$-$$\epsilon_\beta\epsilon_\alpha'$) and 
  symmetric ($\epsilon_\alpha\epsilon_\beta'$$+$$\epsilon_\beta\epsilon_\alpha'$)
  combinations of the incident and outgoing photon polarizations,
  respectively, see Eqs.~(4) and (5). The operators are shown only for 
  $\alpha$=$x$ and $\beta$=$y$
  for the off-diagonal $Q_{\alpha\beta}$ elements, and for $\alpha=z$ for the 
  diagonal $Q_{\alpha\alpha}$ and ${N_\alpha}$; other components follow 
  from symmetry. The parentheses $(-1)$ for $d^5$ and $d^4$ quadrupole
  operators imply an overall minus sign. For $d^3$ configuration with no
  orbital degeneracy, $Q_{\alpha\beta}=0$.}  
  \label{tab:tab1} 
\end{table*}

We make few remarks on the RIXS operators. First, because the resonances at
the $L_3$ and $L_2$ edges involve different intermediate states, the
corresponding operators are different from each other. In particular in the
magnetic scattering channel, they both probe some combinations of \orb\, and
\spin, which are in general not parallel to the total magnetic moment
\mom\,=\,2\spin\,$-$\,\orb\, (minus sign is due to the effective orbital
angular momentum of $t_{2g}$ orbitals\cite{Abragam}). This is made explicit by
denoting the RIXS operator in the magnetic channel by ``${\mathbf N}$'' to
distinguish it from ${\mathbf M}$. Thus, RIXS and INS in general measure two
different quantities even in the magnetic channel unless \orb\, is fully
quenched. It is explicitly confirmed that in the case of $d^3$ systems with
fully quenched \orb, the RIXS operators reduce to pure spin operators. Second,
the ${\mathbf N}$'s for the $L_3$ and $L_2$ edges add up to \orb\, (times some
constant multiplication factor), which is due to a well-known optical sum rule
often used in x-ray magnetic circular dichroism studies\cite{TholePRL92}. 
Similarly, the $Q_{\alpha \beta}$ operators for $L_3$ and $L_2$ sum up to
a pure orbital quadrupoles of corresponding symmetries. This is because summing
up the $L_3$ and $L_2$ edge operators is equivalent to neglecting SOC in the 
core-hole level, and no spin flips are then possible in the RIXS process.      
Third, electron-hole conjugation results (e.g. $d^1\leftrightarrow d^5$) in 
the same operators with an overall minus sign for the quadrupole operators, 
which is not {\it a priori} obvious because of their different intermediate 
multiplet structures. 

As a side remark, we note that if one tries to deduce the diagonal operators
from the off-diagonal  operators (or vice versa) through symmetry
considerations (e.g. $xy$ and $x^2$-$y^2$ are related to each other by $\pi/4$
rotation around $z$-axis), wrong results are obtained with an overall minus
sign; note that $Q_{\alpha \alpha}$ and $Q_{\alpha \beta}$ in Eq.~(4) come
with different signs. In other words, these operators do not rotate like
vectors or dyadics under an arbitrary rotation, because a $t_{2g}$ subsystem
has at most the cubic symmetry.  

In terms of these operators, RIXS cross section in quadrupole and magnetic
channels is expressed as 
\begin{equation}
I_{\omega \mathbf q}\propto\langle R^\dagger R\rangle''_{\omega \mathbf q}
\propto\langle R_Q^\dagger R_Q + R_M^\dagger R_M\rangle''_{\omega \mathbf q}\;.
\end{equation}
Because it is usually difficult to measure x-ray intensities in an absolute
unit unlike in INS, the proportionality constant is unimportant. 

\section{Mapping onto pseudospins}

The RIXS operators we have derived can be applied to any system with
$t_{2g}^n$ electron configuration ($n=1,...,5$). For systems with strong 
SOC (``strong'' implies that SOC dominates over non-cubic crystal fields and
thus orbital moment $L$ remains unquenched), it is more useful to express 
these operators in terms of pseudospin $\tilde{S}$ that spans the low-energy 
manifold of interest. In this section, we provide specific examples of such 
mappings for $d^4$ and $d^5$ configurations, which are particularly relevant 
to ruthenates and iridates.   

\subsection {$d^5$ electronic configuration}

\begin{figure}
\centerline{\includegraphics[width=1\columnwidth,angle=0]{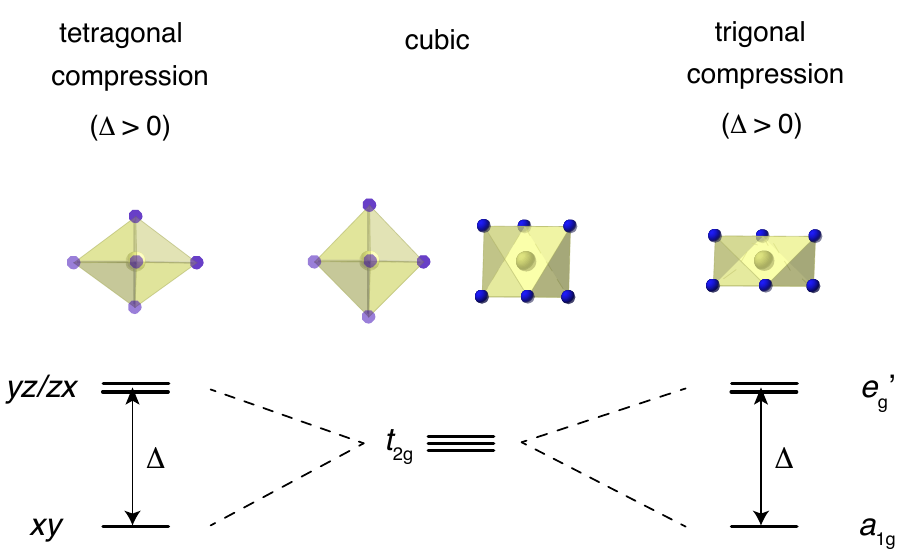}}
\caption{Crystal field splitting in the electron picture. For the case of
  trigonal distortion, $a_{1g}=\frac1{\sqrt3}(xy+yz+zx)$ and   
$e_g'$ = $\bigl\{ \frac1{\sqrt6}(yz+zx-2xy)
\: ;\; \frac1{\sqrt2}(zx-yz) \bigr\}$.}
\end{figure}

The ground state of TM ions with $t_{2g}^5(S=1/2, L=1)$ configuration in
an octahedral crystal field is a Kramers doublet, stabilized by spin-orbit 
coupling. The case is relevant to compounds of Ir$^{4+}$, Rh$^{4+}$,
Co$^{4+}$, Ru$^{3+}$, Os$^{3+}$, etc., ions, and their magnetic properties
can be described in terms of pseudospin $\tilde{S}$=1/2 
\cite{Abragam,Kha04,Khaliullin05,KimPRL08,JackeliPRL09,KimSci09} 
(often referred to as $J_{\textrm{eff}}$\,=\,1/2). 
The $\tilde{S}_z=\pm\tfrac12$ wave functions, denoted as 
$|\tilde\uparrow\rangle$ and $|\tilde\downarrow\rangle$ respectively, 
can be written as 
\begin{align}
|\tilde\uparrow\rangle &=+\sin\theta\, |0,\uparrow\rangle 
-\cos\theta\, |+1,\downarrow\rangle \;, \\ 
|\tilde\downarrow\rangle &=-\sin\theta\, |0,\downarrow\rangle 
+\cos\theta\, |-1,\uparrow\rangle \;, 
\end{align}
\noindent
in the $\ket{L_z,S_z}$ basis with $\ket{L_z=0}$\,=\,$\ket{xy}$ and 
$\ket{Lz=\pm 1}$\,=\,$\mp\frac{1}{\sqrt{2}}$($\ket{yz}$\,$\pm$\,$i\ket{zx})$.  
The quantization axis $z$ is along the axis of tetragonal distortion. From
these expressions, it is easy to see that the wave functions in the limiting 
cases of $\theta\rightarrow0$ ($\theta\rightarrow\pi/2$)
correspond to infinite compression (elongation) of the ligand octahedron
(Fig.~2). In general, the angle $\theta$ parametrizes the distortion through
$\tan2\theta$\,=\,$2\sqrt{2}\zeta/(\zeta+2\Delta_{\textrm{tet}})$, where
$\Delta_{\textrm{tet}}$ and $\zeta$ denote tetragonal crystal field splitting
and SOC, respectively. For example,
$\theta$\,=\,$\frac{1}{2}\arctan{2\sqrt{2}}$ in the cubic limit where
$\Delta_{\textrm{tet}}$\,=\,0, and in the limit $\theta\rightarrow\pi/2$ the
doublet reduces to pure $S$=1/2 states.  
 
Projecting the $d^5$ magnetic ${\mathbf N}$ operators from Table I onto
pseudospin doublet, one finds that 
\begin{equation}
 N_{\alpha}=f_{\alpha}\tilde{S}_{\alpha},\\
\end{equation}
with the following coefficients 
\begin{eqnarray}
f_{x/y}&=&-3\sqrt{2}\sin{2\theta}, \\ 
f_z&=&-2\sqrt{2}(\sin{2\theta}+\sqrt{2}\cos{2\theta})
\end{eqnarray}
for the $L_3$ edge, and
\begin{eqnarray}
f_{x/y}&=&0, \\
f_z&=&-3+\cos{2\theta}+2\sqrt{2}\sin{2\theta}
\end{eqnarray}
for the $L_2$ edge. Note that in the cubic limit ($\cos{2\theta}=1/3$),
Eq.~(13) gives $f_z=0$, so magnetic scattering at the $L_2$ edge vanishes for 
all polarizations\cite{Ame11,Mat16}.   
 
\begin{figure}
\centerline{\includegraphics[width=0.8\columnwidth,angle=0]{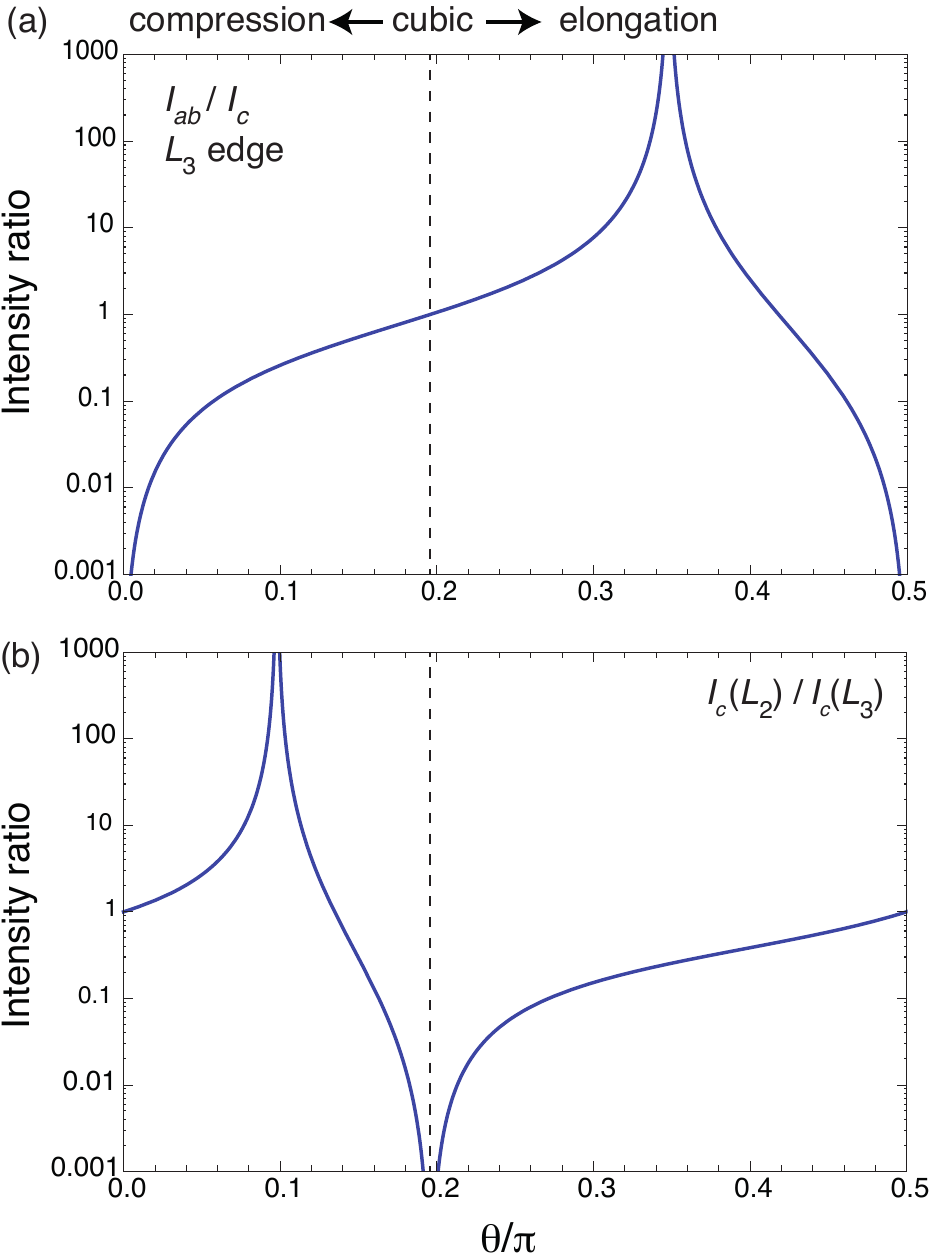}}
\caption{(a) $I_{ab}/I_c$ at the $L_3$ edge. RIXS has different senstivities
  to the in-plane and out-of-plane pseudospin components away from the cubic
  limit at the $L_3$ edge. (b) $I_c$($L_2$)/$I_c$($L_3$) for the out-of-plane
  pseudospin component provides a measure of proximity to the cubic limit. } 
\vspace{-10 pt}
\end{figure}

At the $L_3$ edge, ${\mathbf N}$ becomes isotropic ($f_x$=$f_y$=$f_z$)
in the cubic limit (Fig.~3a) and thus there is a one-to-one correspondence
between the $J$=1/2 dynamics measured by RIXS and $S$=1/2 dynamics measured by
INS, which is behind the surprising similarity between the RIXS spectra of
Sr$_2$IrO$_4$ (Ref. \citenum{KimPRL12}) and INS spectra of La$_2$CuO$_4$
(Ref.~\citenum{BraicovichPRL10}), which are both nearly isotropic (Heisenberg)
antiferromagnets. Away from the cubic limit, RIXS sees different responses for
the $xy$-plane and $z$-axial spin components.  

At the $L_2$ edge, on the other hand, ${\mathbf N}$ is insensitive to the
in-plane spin components. As noted before\cite{LoveseyJPhys11,MorettiPRL14},
this means that resonant x-ray diffraction is blind to magnetic moments lying
in the $xy$ plane regardless of the degree of tetragonal distortion. However,
${\mathbf N}$ has a high sensitivity to tetragonal distortion through $f_z$,
which is identically zero in the cubic limit but rapidly grows 
away from it. Thus, for systems with in-plane moments such as Sr$_2$IrO$_4$,
the proximity to the cubic limit can be measured through the dynamic
out-of-plane fluctuations. For systems with c-axis moments, such as
Sr$_3$Ir$_2$O$_7$ (Ref.~\citenum{JKimPRL12,Boseggia,TakagiPRB12}), the ratio
between the magnetic Bragg peak intensities measured at $L_3$ and $L_2$ edges
are a direct measure of the distortion of the wave functions away from the
cubic limit (Fig.~3b).  

We note that the above discussions also hold for systems with trigonal
distortion ($\Delta_{\textrm{tri}}$) with the redefinition of the 
$\ket{L_z,S_z}$ quantized along the trigonal axis. The corresponding RIXS 
operator ${\mathbf N}$ and explicit expressions for the coefficients 
$f_\alpha$ in the trigonal case can be found in Ref.~\citenum{Cha16}. 
This is applicable to systems such as honeycomb iridates A$_2$IrO$_2$ 
(A=Li,Na). Na$_2$IrO$_3$ is known to have a collinear zig-zag magnetic
structure\cite{ChoiPRL12} with the moment not along the trigonal 
axes\cite{Chun}. We emphasize again that ${\mathbf N}$ is not parallel to
${\mathbf M}$ unless they are both along or perpendicular to the trigonal
axis, and the relation between them is a function of trigonal 
distortion\cite{Cha16}. Thus, an independent measurement of the moment
direction through INS can determine the trigonal distortion through its
comparison to the angle measured by RIXS.   

Finally, the quadrupolar operators $Q_{\alpha\beta}$ vanish identically when
projected to the doublet regardless of distortions, since (pseudo)spin-1/2 
cannot form a quadrupolar moment. The quadrupolar (or higher-multipole) 
scattering is of course symmetry allowed even in the spin-1/2 case, but the 
corresponding RIXS operators should involve at least two neighboring spins in
the scattering process. Such two-site terms are always present in the RIXS
operator expansion\cite{AmentPRB09}, but they are in general weaker and are 
neglected in the present single-ion, local approximation. 

\subsection {$d^4$ electronic configuration}

The case is relevant to compounds of Ru$^{4+}$, Os$^{4+}$, Ir$^{5+}$, etc., ions
in low-spin state with $t_{2g}^4(L=1, S=1)$ configuration. The SOC  
$\frac{\zeta}{2}({\bf L}\cdot{\bf S})$ results in a nonmagnetic ground state 
with total angular momentum $J=0$ (Ref.~\citenum{Abragam}). Magnetic properties of
this class of Mott insulators are governed by collective behavior of
spin-orbit excitons\cite{Khaliullin13}, that is, Van Vleck-type magnetic 
transitions between ground state $J=0$ and excited $J=1$ levels, propagating 
via spin-orbital exchange interactions.  

Figure 4(a) shows the energy level diagram of $t_{2g}^4(L=1, S=1)$ ion as a
function of non-cubic crystal field $\Delta$. As in the $d^5$ case, identical 
results are obtained for tetragonal and trigonal distortions by a suitable 
redefinition of the wave functions. In the cubic limit, the $L=1, S=1$ manifold 
splits by SOC into $J$\,=\,0, 1, and 2 multiplets. At any values of $\Delta$, 
the levels that originate from $J=2$ manifold stay well above the ground
state singlet. The low-energy sector in the cubic limit comprises a ground 
state singlet and $J=1$ triplet. For a non-zero $\Delta$, the triplet splits 
into a singlet and a doublet, derived from $J_z=0$ and $J_z=\pm 1$ states 
correspondingly, either of which merge with the $J$\,=\,0 singlet when 
$\lvert\Delta\lvert$\,$\gg$\,$\zeta$. As a result, the low-energy sector 
contains two quasi-degenerate singlet levels at large negative $\Delta$, while 
a singlet-doublet level system well separated from other levels is formed at 
positive $\Delta$ values. This suggests that the low-energy Hilbert space can 
be described by pseudospin $\tau=1/2$ at large negative $\Delta$, and by 
pseudospin $\tilde{S}=1$ at $\Delta>\zeta$. The pseudospin-1 case, which is 
of particular interest in the context of Ca$_2$RuO$_4$\cite{Jai17}, will be
discussed in detail later in this section. Here we just notice that in the
limit of $\Delta$\,$\rightarrow$\,$\infty$, orbital moment is fully quenched
and the pseudospin 1 becomes identical to pure spin $S=1$, whose magnetism 
is necessarily isotropic (Heisenberg). In contrast, $L_z$ component of 
orbital moment remains unquenched in the pseudospin $\tau=1/2$ limit 
($\Delta$\,$\rightarrow$\,$-\infty$), and an Ising doublet hosting total
magnetic moment with effective $g$-factors $g_c=6$ and $g_{ab}=0$ is formed.  

\begin{figure}
\centerline{\includegraphics[width=0.9\columnwidth,angle=0]{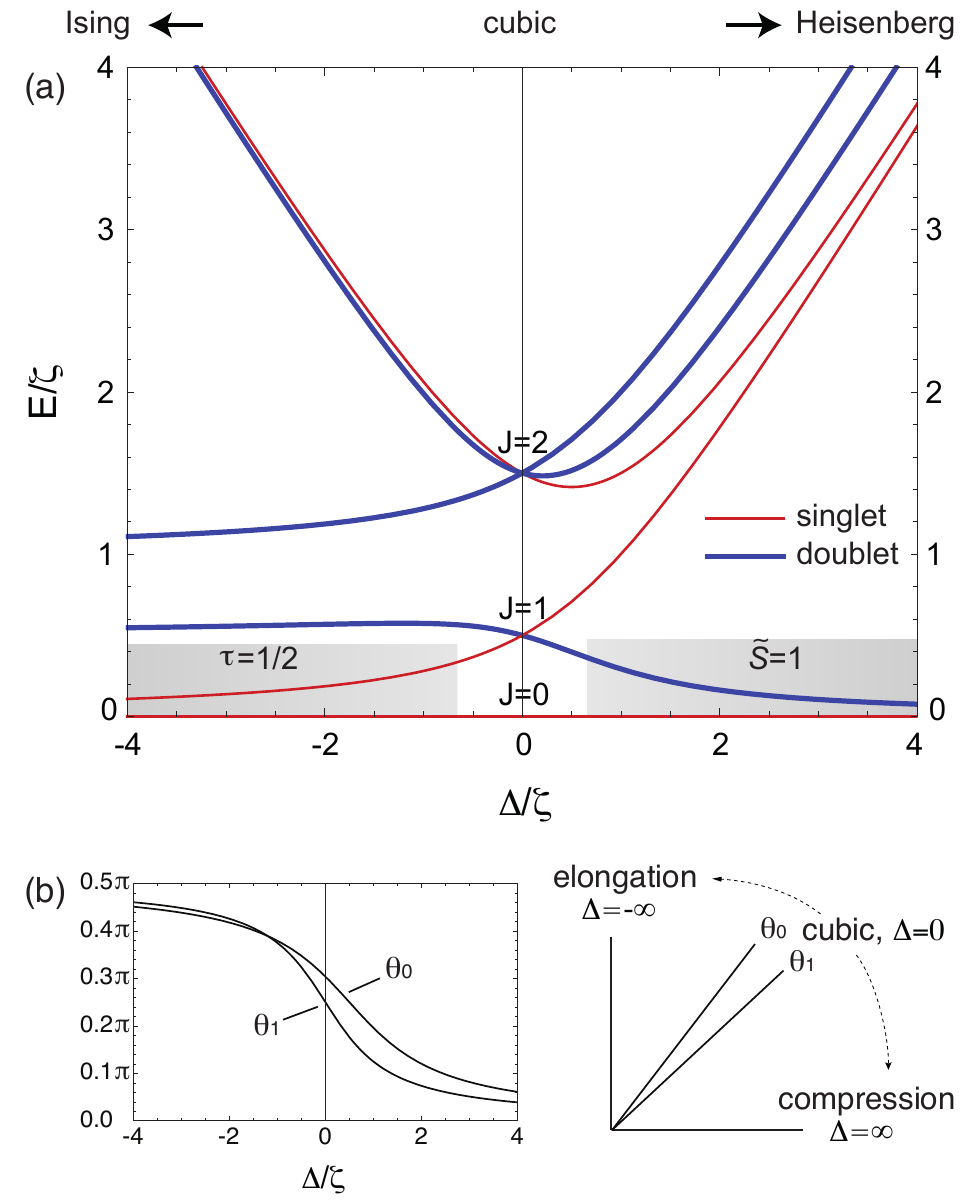}}
\caption{(a) Multiplet energy diagram as a function of non-cubic crystalline 
field $\Delta$, which can be of either tetragonal symmetry $\Delta_{\textrm{tet}}$
or trigonal symmetry $\Delta_{\textrm{tri}}$. Positive $\Delta$ means
compressive distortion. The $J$=0 singlet state energy is taken to be zero.
Thin (thick) lines represent singlet (doublet) states. At large negative 
(positive) values of $\Delta$, ground state singlet 
and the first excited singlet (doublet) levels form a basis for low-energy 
effective $\tau=1/2$ ($\tilde{S}=1$) Hamiltonians. 
(b) The angles $\theta_0$ and $\theta_1$ parametrizing the wave functions in 
Eqs.~(27-29) as functions of $\Delta/\zeta$. In the cubic limit, 
$\theta_0=\arctan{\sqrt{2}}$ and $\theta_1$\,=\,$\pi$/4. In the limit of 
infinite compression (elongation) $\theta_0$\,=\,$\theta_1$\,=\,0 ($\pi/2$).} 
\end{figure}

We now return to the RIXS operators and calculate their matrix elements
within the above single-ion levels. In general (for any $d^n$), one may 
represent the diagonal elements $Q_{\alpha\alpha}$ of the quadrupole 
tensor in terms of cubic harmonics of $A_{1g}$ and ${E_g}$ symmetry:  
\begin{eqnarray}
Q_{r^2}&=& Q_{xx}+Q_{yy}+Q_{zz}\;,\\
Q_{z^2}&=& Q_{xx}+Q_{yy}-2Q_{zz}\;,\\
Q_{x^2-y^2}&=& Q_{xx}-Q_{yy}\;.
\end{eqnarray}
The quadrupole operator $R_Q$ in Eq.~(4) then takes the following form:  
\begin{eqnarray}
R_Q&=& \tfrac{1}{3}(\mathbf{\epsilon\cdot\epsilon'})Q_{r^2}\nonumber\\
&+&\tfrac{1}{6}(\epsilon_x\epsilon'_x+\epsilon_y\epsilon'_y -
2\epsilon_z\epsilon'_z)Q_{z^2}\nonumber\\  
&+&\tfrac{1}{2}(\epsilon_x\epsilon'_x-\epsilon_y\epsilon'_y)Q_{x^2-y^2}\nonumber\\
&-&\tfrac{1}{2}\sum_{\alpha > \beta}(\epsilon_\alpha\epsilon'_\beta -
\epsilon_\beta \epsilon'_\alpha)Q_{\alpha\beta},
\end{eqnarray}
where the last term represents the quadrupoles of $T_{2g}$ symmetry. 
It follows from Table I that $A_{1g}$ component $Q_{r^2}$ is proportional to
the product $({\bf L}\cdot{\bf S})$. For the $d^4$ configuration, it reads as 
$Q_{r^2}=\pm ({\bf L}\cdot{\bf S})$ where upper (lower) sign corresponds to 
the $L_2$ ($L_3$) edge (unessential constant not shown).

We first consider the RIXS matrix elements in the cubic limit. As expected, the
$Q_{\alpha\beta}$ and $\mathbf{N}$ operators allow quadrupole and dipole
transitions with $\Delta J$\,=\,$\pm 2$ and $\Delta J$\,=\,$\pm 1$,
respectively. Thanks to high symmetry of the $J$-wave functions, only few
transitions are allowed. The transition matrix elements in quadrupole sector 
$\bra{J,J_z}Q_{\alpha\beta}\ket{0,0}$ are given as follows:
\begin{eqnarray}
\bra{2,0}Q_{z^2}\ket{0}&=&2\sqrt{2}\cdot
\begin{cases} 
	1, & L_2 \\
	\frac{1}{2}, & L_3
\end{cases}\\
\bra{2,\pm 2}Q_{x^2-y^2}\ket{0}&=&-\frac{2}{\sqrt{3}}\cdot
\begin{cases} 
	1, & L_2 \\
	\frac{1}{2}, & L_3
\end{cases}
\end{eqnarray}
and 
\begin{eqnarray}
\bra{2,0}Q_{xy}\ket{0}&=&\pm \frac{2}{i\sqrt{3}}\cdot
\begin{cases} 
	1, & L_2 \\
	\frac{1}{2}, & L_3
\end{cases}\\
\bra{2,\pm 1}Q_{yz}\ket{0}&=&-\frac{2}{i\sqrt{3}}\cdot
\begin{cases} 
	1, & L_2 \\
	\frac{1}{2}, & L_3
\end{cases}\\
\bra{2,\pm 1}Q_{zx}\ket{0}&=&\mp \frac{2}{\sqrt{3}}\cdot
\begin{cases} 
	1, & L_2 \\
	\frac{1}{2}. & L_3
\end{cases}
\end{eqnarray}
For the magnetic scattering ${\mathbf N}$ operator, the matrix elements for 
transitions from $J=0$ ground state to $J=1$ level are concisely written as 
\begin{eqnarray}
\bra{1,0}N_z\ket{0}&=&\bra{1,\pm 1}\frac{1}{\sqrt{2}}(N_x \pm iN_y)\ket{0}
\nonumber\\ 
&=&-i{\sqrt{6}}\cdot
\begin{cases} 
	0, & L_2 \\
	1. & L_3
\end{cases}
\end{eqnarray}
Since the ground state $J$\,=\,0 is a nonmagnetic singlet, the magnetism 
necessarily involves $J=1$ states\cite{Khaliullin13,Meetei15}. It is 
interesting to note that the ``excitonic'' magnetism\cite{Khaliullin13} arising
from Van Vleck-type transitions between the $J$\,=\,0 and $J$\,=\,1 manifolds 
can only be probed at the $L_3$ edge, and thus this selection rules serve as 
a means to differentiate between magnetic moments derived from spin-orbit 
excitons and that from conventional origins, {\it e.g.} $S$\,=\,1 moments 
with quenched orbital moments. We also note that transitions within the
excited states are also ``edge-selective''; e.g., transitions within the 
$J=1$ manifold is allowed only at the $L_2$ edge:
\begin{eqnarray}
\pm\bra{1,\pm 1}N_z\ket{1,\pm 1}&=&\bra{1,\pm 1}\frac{1}{\sqrt{2}}
(N_x \pm i N_y)\ket{1,0}\nonumber\\
&=&\frac{3}{2}\cdot
\begin{cases} 
	1, & L_2 \\
	0. & L_3
\end{cases}
\end{eqnarray}
These selection rules are summarized in Fig.~5. For completeness, the figure 
includes the allowed transitions also within the excited states (dashed lines).

\begin{figure}
\centerline{\includegraphics[width=0.9\columnwidth,angle=0]{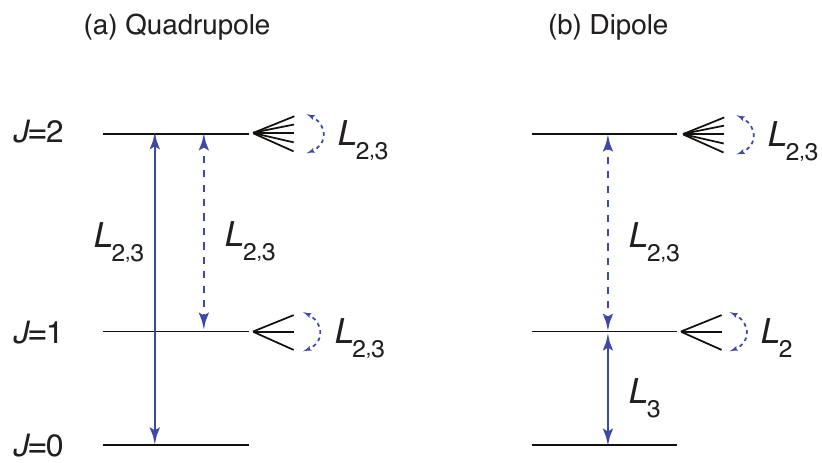}}
\caption{Allowed transitions for the (a) quadrupole and (b) dipole operators 
in case of cubic symmetry. Transitions within the excited states $J=1$ 
and $J=2$ are indicated by dashed lines; these processes are not important 
unless $J=1, 2$ states are strongly mixed in the many-body ground state.} 
\end{figure}

Away from the ideal cubic limit, spin-orbit wave functions and hence the above 
selection rules are gradually modified. In the extreme cases of  
$\Delta$\,$\rightarrow$\,$-\infty$ and $\Delta$\,$\rightarrow$\,$\infty$, 
corresponding to Ising $\tau=1/2$ and Heisenberg $S=1$ limits, the RIXS 
operators within the respective low-energy sectors read as follows ($R_Q=0$): 
\begin{eqnarray}
-R_M(L_2)=\tfrac{1}{4}R_M(L_3)
&=&(\epsilon_x \epsilon_y'-\epsilon_y \epsilon_x')\tau_z\;, \\
R_M(L_2)=-R_M(L_3)&=&
(\epsilon_y \epsilon_z'-\epsilon_z \epsilon_y')S_x \nonumber \\
&+&(\epsilon_z \epsilon_x'-\epsilon_x \epsilon_z')S_y\;.
\end{eqnarray}
These expressions tell that at negative $\Delta$ values, the $L_3$ edge is
still dominant in the magnetic RIXS process, while the $L_2$ edge becomes of 
comparable strength at large $\Delta>0$. It is also noticed that the two
limits have an opposite (out-of-plane versus in-plane) polarization
dependences. 

\begin{table*}
\newcolumntype{R}{>{\raggedleft\arraybackslash}X}

	\begin{tabularx}{1.0\textwidth}{|l|c|c||c|c|}
		\multicolumn{1}{c}{}& 
		\multicolumn{1}{c}{$L_3$ edge} &
		\multicolumn{1}{c}{cubic}& 
		\multicolumn{1}{c}{$L_2$ edge}&
		\multicolumn{1}{c}{cubic}\\	 
		\cline{1-5} 
		$a_{xy}$ &
                $-2\sin{\theta_0}(\sqrt{2}\cos{\theta_0}+\sin{\theta_0})+
2\sin{\theta_1}(\sin{\theta_1}+\cos{\theta_1})$
                &$-2/3$
                &$\sin{\theta_0}(2\sqrt{2}\cos{\theta_0}-\sin{\theta_0})-
\sin{\theta_1}(2\cos{\theta_1}-\sin{\theta_1})
                $& 1/6\\  
		\cline{1-5}
		$a_z$ & $-2(2\sin^2{\theta_1}-\sin^2{\theta_0})$ & $-2/3$  
& $2(2\sin^2{\theta_0}-\sin^2{\theta_1})$ & 5/3 \\
		\cline{1-5} 
		$b_{xy}$  & $2\sin{\theta_1}(\sin{\theta_1}-\cos{\theta_1})$ 
& 0& $\sin{\theta_1}(\sin{\theta_1}+2\cos{\theta_1})$ & 3/2 \\
		\cline{1-5}
		$b_{z}$ &
                $\frac{1}{\sqrt{2}}\sin{\theta_0}(\cos{\theta_1}+
\sin{\theta_1})-2\cos{\theta_0}\sin{\theta_1}$
                & 0 &
                $\sqrt{2}\sin{\theta_0}(\cos{\theta_1}-\frac{1}{2}
\sin{\theta_1})-\cos{\theta_0}\sin{\theta_1}$
                & 0 \\ 
		\cline{1-5}
		$c_{xy}$ &
                $-(2\cos{\theta_0}+\frac{1}{\sqrt{2}}\sin{\theta_0})
(\cos{\theta_1}+\sin{\theta_1})$
                & $-\sqrt{6}$ &
                $\cos{\theta_0}(2\cos{\theta_1}-\sin{\theta_1})-
\sqrt{2}\sin{\theta_0}(\cos{\theta_1}-\frac{1}{2}\sin{\theta_1})$
                & 0 \\ 
		\cline{1-5}
		$c_z$ & $-2\sin{\theta_1}(\cos{\theta_1}-
\sin{\theta_1})$ & 0 & $\sin{\theta_1}(\sin{\theta_1}+2\cos{\theta_1})$ & 3/2 \\
		\cline{1-5}
	\end{tabularx}
\caption{The coefficients $a_\gamma$, $b_\gamma$, and $c_\gamma$ in Eqs.~(30)
         and (31) as functions of the pseudospin wave function angles 
         $\theta_0$, $\theta_1$ . The values in the cubic limit are also given.}
\label{tab:tab2} 
\end{table*}

Having in mind the Mott insulator Ca$_2$RuO$_4$, which has been recently 
confirmed\cite{Jai17,Sou17} to host spin-orbit excitonic 
magnetism\cite{Khaliullin13}, we now consider the case of compressive 
distortion ($\Delta>0$) in a greater detail. Figure 4(a) shows that already 
at $\Delta$\,$\sim$\,$\zeta$, a singlet split off from the $J=1$ triplet 
quickly goes high in energy and thus the low-energy physics can be well 
approximated as a three-state (singlet plus doublet) system, i.e. by an 
effective $\tilde{S}=1$\cite{Jai17,Sou17}. In terms of $\ket{L_z,S_z}$ basis, 
the pseudospin $\tilde{S}_z$ states are expressed as 
\begin{eqnarray}
 \ket{+\tilde{1}}&=&\cos{\theta_1}\ket{0,1}-\sin{\theta_1}\ket{1,0}, \\
 \ket{\tilde{0}}&=&\cos{\theta_0}\ket{0,0}- 
\frac{\sin{\theta_0}}{\sqrt{2}} (\ket{-1,1}+\ket{1,-1}), \\ 
 \ket{-\tilde{1}}&=&\cos{\theta_1}\ket{0,-1}-\sin{\theta_1}\ket{-1,0},
\end{eqnarray}
where the two angles $\theta_0$\,$\leq$\,$\pi/2$ and
$\theta_1$\,$\leq$\,$\pi/2$ parametrize the distortion through
$\tan\theta_0$\,=\,$(\sqrt{9\zeta^2-4\Delta\zeta+4\Delta^2}+
\zeta-2\Delta$)/$2\sqrt{2}\zeta$
and $\tan\theta_1$\,=\,$\zeta/(\Delta$+$\sqrt{\Delta^2+\zeta^2}$). In the
cubic limit, $\theta_0=\arctan{\sqrt{2}}$ and $\theta_1=\pi$/4. Figure 4(b) 
shows $\theta_0$ and $\theta_1$ at arbitrary values of $\Delta/\zeta$. 

\begin{figure}[b]
\centerline{\includegraphics[width=1\columnwidth,angle=0]{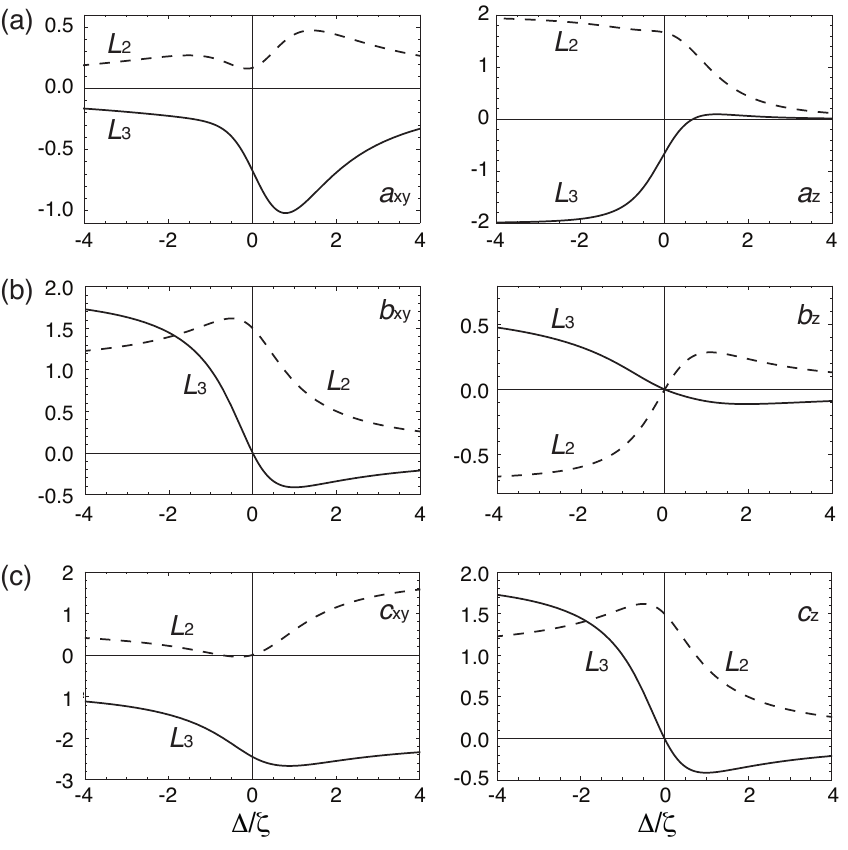}}
\caption{The coefficients (a) $a_\gamma$, (b) $b_\gamma$, and (c) $c_\gamma$ 
         in Eqs.~(30) and (31) as functions of $\Delta/\zeta$.} 
\end{figure}

Using the results of Table I for $d^4$ quadrupole and magnetic operators, 
we evaluate their matrix elements within the above $\tilde{S}=1$ manifold.  
The results are then expressed in terms of pseudospin operators. This gives 
an effective RIXS operator $R_Q+iR_M$ with 
\begin{eqnarray}
R_{Q}&=&\tfrac{1}{2}[a_{xy}(\epsilon_x \epsilon_x'+\epsilon_y \epsilon_y')+
a_z \epsilon_z \epsilon_z' ]\tilde{S}_z^2\nonumber\\
 &+&\tfrac{1}{2}b_{xy}[(\epsilon_x \epsilon_x'-\epsilon_y \epsilon_y') 
(\tilde{S}_y^2-\tilde{S}_x^2)\nonumber\\
&+& (\epsilon_x\epsilon_y'+\epsilon_y \epsilon_x')
(\tilde{S}_x\tilde{S}_y+\tilde{S}_y\tilde{S}_x)]\nonumber\\
&+&\tfrac{1}{2}b_z[(\epsilon_y \epsilon_z'+\epsilon_z \epsilon_y')
(\tilde{S}_y\tilde{S}_z+\tilde{S}_z\tilde{S}_y)\nonumber\\
&+& (\epsilon_z \epsilon_x'+\epsilon_x \epsilon_z')
(\tilde{S}_z\tilde{S}_x+\tilde{S}_x\tilde{S}_z)]\;, \\
R_{M}&=&\tfrac{1}{2}c_{xy}[(\epsilon_y\epsilon_z'-\epsilon_z\epsilon_y')
\tilde{S}_x+(\epsilon_z\epsilon_x'-\epsilon_x\epsilon_z')\tilde{S}_y]\nonumber\\
 &+&\tfrac{1}{2}c_z(\epsilon_x\epsilon_y'-\epsilon_y\epsilon_x')
\tilde{S}_{z}\;.
\end{eqnarray}
The coefficients $a_\gamma$, $b_\gamma$, and $c_\gamma$ are as summarized 
in Table II and plotted in Fig.~6. In the limit of 
$\Delta$\,$\rightarrow$\,$\infty$, only $c_{xy}$ is finite; all other 
terms vanish, while $c_{xy}(L_2)$\,=$-c_{xy}(L_3)$. 

As in the case of $d^5$ system, RIXS is sensitive to magnetic dipole moments
through the $\mathbf{N}$ operator, and relative intensities of the $L_3$ and 
$L_2$ edges may help to quantify the $\Delta/\zeta$ ratio. A distinct 
difference from the $d^5$ case is that RIXS is sensitive to quadrupole moments 
which are expressed in terms of $\tilde{S}$\,=\,1 operators in Eq.~(30). 

The effective $\tilde{S}$\,=\,1 RIXS operator (30-31) and its parameters in 
Table II should be useful for quantitative analysis of RIXS experiments in 
$d^4$ systems including Ca$_2$RuO$_4$, where combined action of SOC and crystal
fields results in a singlet-doublet level structure as shown in Fig.~4(a). 
We note that Eqs.~(30-31) and Table II remain valid for arbitrary values of 
$\Delta$; however, they only concern the transitions within the singlet-doublet
subsystem. At small and/or negative $\Delta$ values, transition to the singlet 
derived from $J_z=0$ state become relevant and have to be included in the 
low-energy RIXS operator.   

Magnetic order in compounds based on Van Vleck-type $d^4$ ions is due to 
Bose-Einstein condensation of the higher lying magnetic 
states\cite{Khaliullin13}, and collective excitations in the ordered state 
comprise, in addition to conventional magnons, the amplitude (Higgs) mode. 
The latter has recently been detected by neutron\cite{Jai17} and 
Raman\cite{Sou17} scattering studies, and the present work suggests that 
the Higgs mode can be directly accessed by the RIXS. Indeed, the first term 
in $R_Q$ of Eq.~(30), which is proportional to $\tilde{S}_z^2$ should couple 
to the length fluctuations of the magnetic order parameter in Ca$_2$RuO$_4$. 
These results show also that RIXS is useful for detecting a spin quadrupolar 
(nematic) order\cite{Podolsky05} and its associated collective excitations.  

\section{Summary}
Despite the fact that RIXS has in the recent years become a very popular tool
for probing magnetism, the quantities measured by RIXS have not been known
precisely particularly for 4$d$ and 5$d$ TM compounds, which generally have
unquenched orbital moments in addition to spin moments. In this paper, we have
derived general expressions for RIXS operators for $t_{2g}$ orbital systems;
the RIXS operators are expressed in terms of $\mathbf{L}$ and $\mathbf{S}$
under the fast collision approximation, which is valid for Mott insulators
where spin and orbital energy scales are lower than core-hole inverse
lifetimes.

In 4$d$ and 5$d$ TM compounds, spin and orbital moments are coupled through
strong intra-ionic spin-orbit coupling, and behave as one composite object
that can be represented as pseudospins in certain limits. The RIXS operators
are then more concisely expressed in terms of pseudospins, which offer more
insights into the physics they realize, through mapping onto spin-only
Hamiltonians for which a large body of theoretical studies are available. We
have discussed the cases for $\tilde{S}$\,=\,1/2 and $\tilde{S}$\,=\,1
realized in some iridates and ruthenates, respectively. For iridates, our
approach makes the physical reason behind the surprising similarities in the
spin excitation spectra between iridates and cuprates more transparent. For
ruthenates, we have shown that RIXS is capable of probing quadrupole moments
in addition to dipole moments. For systems that lack static dipole moments,
RIXS thus becomes a useful tool for detecting (pseudo)spin nematic order,
which for pure-spin systems has been very challenging.   

The RIXS operators documented in this paper can be useful for quantitative
simulations and their comparisons with experimental spectra of a broad class
of TM compounds with $t_{2g}$ orbital degeneracy. 

\acknowledgements{We would like to thank M. Minola for useful comments. We acknowledge support by the European Research
Council under Advanced Grant 669550 (Com4Com), and by IBS-R014-A2.}

\bibliography{bib.bib}

\end{document}